\documentclass[twocolumn,10pt,pra,superscriptaddress]{revtex4-1}
\usepackage{amsfonts,graphicx}
\usepackage{xcolor}
\usepackage{times}
\usepackage{hyperref}
\hypersetup{colorlinks=true, linkcolor=red, citecolor=blue}

\begin{document}

\title{High fidelity quantum state transfer in electromechanical systems
 with intermediate coupling}

\author{Jian Zhou}
\affiliation{Laboratory of Quantum Engineering and Quantum Materials, and
School of Physics\\ and Telecommunication Engineering,  South China Normal University, Guangzhou 510006, China}
\affiliation{Anhui Xinhua University, Hefei, 230088, China}

\author{Yong Hu}
\affiliation{Department of Physics and Center of Theoretical and Computational Physics,  The University of Hong Kong, Pokfulam Road, Hong Kong, China}
\affiliation{School of Physics, Huazhong University of Science and Technology, Wuhan 430074, China}

\author{Zhang-qi Yin}
\affiliation{Center for Quantum Information, Institute for Interdisciplinary Information Sciences, Tsinghua University, Beijing 100084, China}

\author{Z. D. Wang}
\affiliation{Department of Physics and Center of Theoretical and Computational Physics, The University of Hong Kong, Pokfulam Road, Hong Kong, China}

\author{Shi-Liang Zhu} \email{slzhunju@163.com}
\affiliation{National Laboratory of Solid State Microstructure and Department of Physics, Nanjing University, Nanjing 210093, China}

\author{Zheng-Yuan Xue} \email{zyxue@scnu.edu.cn}
\affiliation{Laboratory of Quantum Engineering and Quantum Materials, and School of Physics\\ and Telecommunication Engineering,  South China Normal University, Guangzhou 510006, China}
\affiliation{Department of Physics and Center of Theoretical and Computational Physics, The University of Hong Kong, Pokfulam Road, Hong Kong, China}

\date{\today}

\begin{abstract}
\textbf{Hybrid quantum systems usually consist of two or more subsystems, which may take the advantages of the different systems. Recently, the hybrid system consisting of circuit electromechanical subsystems have attracted great attention due to its advanced fabrication and scalable integrated photonic circuit techniques. Here, we propose a scheme for  high fidelity quantum state transfer between a superconducting qubit and a nitrogen-vacancy center in diamond,  which are coupled to a  superconducting transmission-line resonator with coupling strength $g_1$ and  a nanomechanical resonator with coupling strength $g_2$, respectively. Meanwhile, the two resonators are parametrically coupled  with coupling strength $J$. The system dynamics, including the decoherence effects, is numerical investigated. It is found that both the small ($J \ll \{g_1, g_2\}$) and large ($J \gg \{g_1, g_2\}$) coupling regimes of this hybrid system can not support high fidelity quantum state transfer before significant technique advances. However, in the intermediate coupling regime ($J \sim g_1 \sim g_2$),  in contrast to a conventional wisdom,  high fidelity quantum information transfer can be implemented, providing a promising route towards high fidelity quantum state transfer in similar coupled resonators systems.}
\end{abstract}


\maketitle

A rapid processor and reliable memory are indispensable components for building quantum computers. Recent progress  in mesoscopic objects, e.g., superconducting qubits, shows that this kind of artificial systems can couple strongly to electromagnetic field and enable fast  quantum logic gates \cite{rjs}. However, they possess relatively short coherence time. On the other hand, microscopic systems, e.g., nitrogen-vacancy (NV) centers in  diamond
\cite{Balasubramanian2009}, naturally have rather long coherence time. Therefore, it is attractive to combine the two types of systems by employing their hybrid architecture~\cite{hybrid}. In order to take the advantages of both, quantum state transfer (QST) between the two sides is crucial. To this end, some strategies have been proposed with hybrid systems \cite{Tordrup2008,Petrosyan2009,Yang2011b,Kubo2011,nvs,nvs2,nvs3},
where a superconducting qubit acts as a processing unit and the other ones serve as memory units.

There are generally two kinds of hybridity: coupling the two elements via a quantum bus~\cite{Kubo2011} or directly~\cite{nvs}. First, a superconducting qubit can be coherently coupled to an ensemble of NV centers in a diamond via a superconducting
transmission-line resonator, which acts as a quantum bus~\cite{Yang2011b,Kubo2011}. However, the NV center ensemble usually has relatively much shorter coherence time comparing with single NV center and the coupling of single NV center to the
transmission-line resonator  is too weak to be usable, usually under 0.1 kHz  \cite{Kubo2011}.   Meanwhile, a superconducting flux qubit can be directly coupled to an NV center with a coupling strength on the order of 10 kHZ ~\cite{nvs,nvs2,nvs3}.  But, such schemes are difficult to  extend to couple long distance qubits as the  superconducting qubit can not be too large. Therefore, to be potentially extended to quantum networks \cite{Cirac1997}, many QST schemes have been investigated in systems  consisting of two coupled resonators
\cite{suncp,Li2009a,yin2009,sing,Yin2007,zhang2010a,Yang1,Yang2,ks1,ks2}.
Typically, an  NV center and a superconducting qubit can be coupled
to a nanomechanical resonator \cite{Rabl2009,Rabl2010,Arcizet2011}
and a transmission-line  resonator \cite{rjs}, respectively. On
the other hand, the two different resonators can be parametrically
coupled to form an electromechanical architecture \cite{em}.  With
rapid experimental progress \cite{em,Zhou2013,Pirkkalainen2013,Palomaki2013}, this typical system can have many potential applications for hybrid quantum information processing~\cite{Yin2007,ks1,ks2} and entanglement \cite{em1,em2,em3}.

In this paper, we propose a scheme to realize  high fidelity QST between a superconducting qubit and an NV center,  which provides the possibility of storing a state of the superconducting qubit into the spin-based quantum memory via the electromechanical system. Previous works in similar coupled resonator systems, such
as Refs. \cite{zhang2010a,Yang1,Yang2,ks1,ks2}, were mainly focusing on the large inter-resonator  coupling regime (we will specify different regimes later), where analytical result can be obtained \cite{Yang2}. However,  after taking the decoherence effects into consideration, we investigate the QST dynamics numerically and find that both the small and large coupling regimes of this system can not support high fidelity QST with current technology.  The main result of present work is the comprehensive study of QST in the intermediate coupling regime, which is lacking in previous studies  \cite{zhang2010a,Yang1,Yang2,ks1,ks2}. Our motivation also relies on the fact that the intermediate coupling regime can be realized more easily as it requires smaller inter-resonator  coupling comparing with the large coupling regime, and thus our result is useful and can be tested in a near future experiment.  Interestingly, the main finding here is that fast and high fidelity QST is possible in the intermediate coupling regime, even better than that of the large coupling regime. The physics behind this surprising result is that the decoherene has less effect in the intermediate coupling regime since the time of QST is shorter than that of both the small and large coupling regimes.  Therefore, our result provides a promising route towards high fidelity QST in similar systems with coupled  resonators.

\bigskip

\noindent\textbf{Results}

\noindent\textbf{The hybrid system.}
The hybrid system consists of a superconducting qubit (NV center)
coupled to a transmission-line  (nanomechanical)   resonator and
the two resonators are  parametrically coupled. Our final goal is to
transfer the quantum state of a superconducting qubit, the
processor, to the NV center, the quantum memory, via the two
coupled resonators. Since the QST process interchanges the
information between the two subsystems, the main results we
obtained in the following should also be valid for the reverse
transfer, i.e., transfer the state of the NV center to the
superconducting qubit. For convenience, hereafter, the subscript
$1(2)$ stands for the superconducting qubit (NV center) and $a
(b)$ for the transmission-line  (nanomechanical)  resonator,
respectively. A superconducting qubit  can be treated as a
two-level system and the transmission-line  resonator can be
modeled as a single Bosonic mode \cite{blias}: $H_{\text{a}} =
\omega_{\text{a}} a^\dag a$, where we  assume $\hbar = 1$
hereafter, $\omega_{\text{a}} $ and $a$ ($a^\dag$) are the mode
frequency and the annihilation (creation) operator of the
resonator, respectively. The superconducting
qubit can be capacitively coupled to the transmission-line
resonator and  the interaction Hamiltonian can be written as
\cite{blias,zhu}
\begin{eqnarray} \label{h1}
H_{a,1} = g_{1}(\sigma_1^- a^\dag + \sigma_1^+ a),
\end{eqnarray}
where $\sigma^\pm = (\sigma^x \pm i\sigma^y)/2$ are the
raising/lowering operators of the superconducting qubit and  the
coupling strength $g_{1}$  can be tuned by an external magnetic
flux of the qubit loop \cite{blias,zhu}.

Meanwhile, the ground state of a negatively charged NV center is spin triplet with $m_s=0$ and $m_s=\pm 1$, and there is a zero-field splitting between them due to the spin-spin interaction. The two states $|\pm 1\rangle$ are coupled to the  $|0\rangle$ state by different polarized driven fields with the same detuning $\Delta>0$ and strength $\Omega$ \cite{du}.   In a frame rotating with the driven frequencies, the Hamiltonian of the NV center can be written as
$$H_{\text{NVC}}=\sum_{i=\pm 1} \left[\Delta |i\rangle\langle i|+{1
\over 2} \Omega(|0\rangle\langle i|+|i\rangle\langle 0|) \right],$$
which couples $|0\rangle$ to the bright state  $|b\rangle=(|-1\rangle+|1\rangle)/\sqrt{2}$, while leaves the dark state $|d\rangle=(|-1\rangle-|1\rangle)/\sqrt{2}$ uncoupled. The eigenstates of the NV center are $|d\rangle$, $|m\rangle=\cos\vartheta|0\rangle-\sin\vartheta|b\rangle$, and $|e\rangle=\cos\vartheta|b\rangle+\sin\vartheta|0\rangle$ with $\tan(2\vartheta)=\sqrt{2}\Omega/\Delta$ and the corresponding eigenvalues are labeled as $\omega_{d,m,e}$. The Hamiltonian of a nanomechanical  resonator can be written as $H_b = \omega_{_{\text{b}}} b^\dag b$ with $\omega_b$ and $b$ ($b^\dag$)
being the mode frequency and the annihilation (creation) operator of the nanomechanical  resonator, respectively. A magnetic tip is attached to the nanomechanical  resonator at a distance $h\sim 25$ nm above a NV center, and thus creates a strong coupling between the two elements. The mechanical resonator can be a doubly clamped beam or a cantilever, and the magnetic tip can be attached at the position of an antinode and the unclamped end, respectively. The superconducting qubit is far away from the tip (comparing with $h$), and thus feels negligible magnetic field. When $\omega_b=\omega_{dm}=\omega_d-\omega_m$, the  interaction between the NV center and the nanomechanical  resonator reads
\cite{Rabl2009}
\begin{eqnarray}\label{h2}
H_{b,2}= g_{2}(b \sigma_2^\dag   + b^\dag \sigma_2^-),
\end{eqnarray}
where $\sigma_2^-=|m\rangle\langle d|$,  $\sigma_2^+= (\sigma_2^-)^\dag$, and $g_2\ (\ll \omega_b)$ is the coupling strength.

Furthermore, we consider a capacitive coupling  between the microwave transmission-line  and the  nanomechanical resonator \cite{em,Palomaki2013}.  The mechanical resonator is driven by an ac voltage, which matches the energy difference between the electrical and mechanical systems, and thus couples the two. Then, an effective linear coupling between the two resonators can be induced, and the effective electromechanical interaction Hamiltonian can be written as
\begin{eqnarray}\label{Hab}
H_{a,b}=  J(a b^\dag + a^\dag b),
\end{eqnarray}
where  the  coupling strength $J$ determines the speed of  the energy transfer between the two resonators. The  nanomechanical resonator here can be a doubly clamped beam  \cite{Palomaki2013} or a metallic membrane \cite{em}. However, the direct capacitive coupling  strength of the microwave resonator and the mechanical beam is usually very small in typical experiment \cite{Palomaki2013}. Fortunately, one can replace the coupling capacitor by a superconducting qubit, which can greatly enhance the coupling strength \cite{enhance}. Alternatively, one can use a metallic membrane to serve as one of the two parallel metal plates of the coupling  capacitor  \cite{em}, where one can readily obtain $J\sim g_2$. In this case, the coupling between the membrane and the NV center can be induced by strain \cite{strain}, where strong coupling can be achieved. Therefore, the total Hamiltonian for the hybrid system is given as
\begin{eqnarray}\label{Ht}
H_{t}= H_{\text{a,1}}  + H_{\text{b,2}} + H_{\text{a,b}}.
\end{eqnarray}
The competition of different couplings of this Hamiltonian will yield three coupling regimes.  As $J$ determines the time scale of the systematic dynamics, we will call $J\gg \{g_1, g_2\}$, $J\ll  \{g_1, g_2\}$, and  $J\sim g_1 \sim g_2$ as large, small, and intermediate  (inter-resonator) coupling regime, respectively.
In addition, to get coherent dynamics, the coupled systems described in Hamiltonian in Eqs. (\ref{h1}) and (\ref{h2}) need also to be operated in the strong coupling regime.

\bigskip

\noindent\textbf{The quantum state transfer.}
We now study the QST from a superconducting qubit to a NV center.  We consider the zero- and one-excitation subspaces, i.e., spanned by the basis of $\{|0m\rangle_{12}|10\rangle_{ab}$, $|0m\rangle_{12}|01\rangle_{ab}$,
$|1m\rangle_{12}|00\rangle_{ab}$, $|0d\rangle_{12}|00\rangle_{ab}\}$. The basis vectors and their corresponding populations  are labeled as $|\phi_n\rangle$  and
$P_n$ ($n=1, 2, 3, 4$), respectively. For an initial state
$|\psi\rangle_i=|\psi_1\rangle |m\rangle_2 |00\rangle_{ab}$ with $|\psi_1\rangle =\cos \theta |1\rangle_1 + \sin \theta |0\rangle_1$, the goal of the QST is
to obtain a final state of $|\psi\rangle_f=|0\rangle_1(\cos \theta
|d\rangle_2 + \sin \theta |m\rangle_2)|00\rangle_{ab}$. Note  that  $|0\rangle_1 |m\rangle_2|00\rangle_{ab}$ will not evolve under the act of  Hamiltonian  (\ref{Ht}) and the QST completes when the excitation initially in the
superconducting qubit ($P_3=1$) transfers to the NV center ($P_4$). We estimate the performance of our scheme by the conditional fidelity  defined by  $F=\langle\psi_1|\rho_a|\psi_1\rangle$, with $\rho_a$ being the reduced  density
matrix of the NV center from the final state, under decoherence rates $\xi$ using the Lindblad master equation, see the method section for details.

\begin{figure}[tb]\centering
\includegraphics[width=7cm]{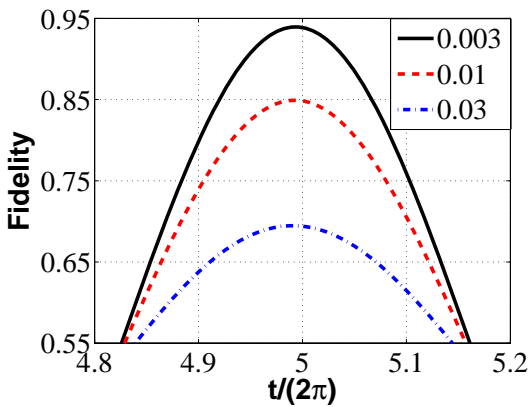}
\caption{Fidelity  as a function of the reduced time of $t/(2\pi)$ in
the small coupling regime with $J=0.1$ for different $\xi$, other parameters are $\theta=\pi/4$, $\zeta=0.001$,  and $g_1=g_2=1$. }
\label{weak}
\end{figure}

\bigskip

\noindent\textbf{The small coupling regime.}
In this regime, the dynamics can be understood as two weak  coupled subsystems with
each having a  Jaynes-Cummings type interaction. The dynamics of the system is characterized by a fast oscillation behavior with frequency $g=g_1=g_2$ accompanied by a slow oscillation envelope with frequency $J/2$. Choosing
$\theta=\pi/4$, the fidelities are plotted in Fig. \ref{weak} with different $\xi$. This transfer is completed at instants $T_w = (2k + 1)\pi/J $ with $k$ being a nonnegative integer. However, the maximum fidelity  with $\xi=0.03$ is less than 70\%. The relatively large infidelity is mainly due to the decoherence effect since the transfer time is very long because of the very small $J$. To obtain a fidelity about 95\%, one needs to reduce the decay rates to $\xi/10$, which is very
challenging. Moreover, it is worth to note that if the symmetric
coupling situation $g_1 = g_2$ is not satisfied, the deviation
brings detrimental influence to the QST: the population maximum
will decrease rapidly with the increase of the deviation denoted
by $|g_2-g_1|/g_2$. Furthermore, the fidelity  is too small, which means
it is impossible for QST with asymmetric coupling in this regime.

\begin{figure}[tb]\centering
\includegraphics[width=6cm]{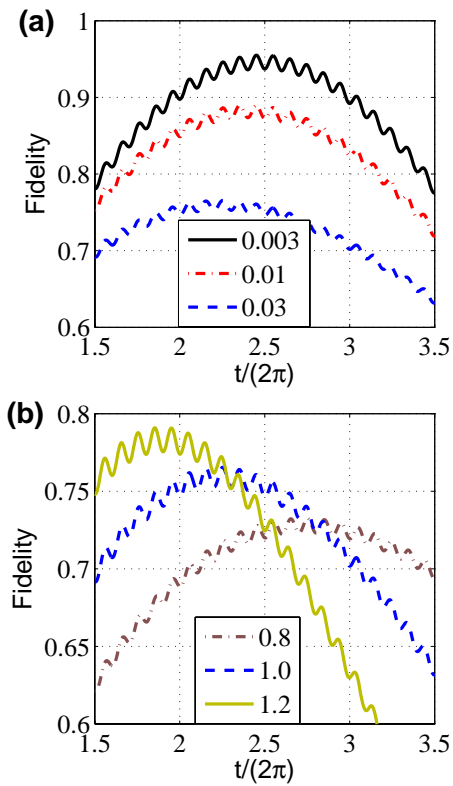}
\caption{Fidelity as a function of the reduced time of $t/(2\pi)$  in the large coupling regime with $J=10$,  other parameters are
$\theta=\pi/4$ and $\zeta=0.001$. (a) For $g_1=g_2=1$ with  different $\xi$. (b) For $g_2=1$ and $\xi=0.03$ with  different $g_1$.}
\label{strong}
\end{figure}

\bigskip

\noindent\textbf{The large coupling regime.}
In this regime, the system can be considered as two  subsystems connected by large photon hopping. The relatively large coupling between the two resonators
services as a fast channel to transfer the energy residing in the
two qubits. The system dynamics can be characterized by  two
distinct oscillating frequencies \cite{Yang2}, fast oscillation with frequency
$J$ and slow oscillation with frequency $g^2/J$. QST
can be completed at instants $T_s = (k + 1/2)\pi J/g^2$ with $k$
being a nonnegative integer. The performance of this regime is
better than that of the small coupling regime, as shown in  Fig.
\ref{strong}(a). The maximum  fidelity of the transfer process is
about 75\% for present experimental parameter of $\xi=0.03$. To
obtain a fidelity above 90\% and 95\%, one needs to reduce the
decay of the system to $\xi/3$ and $\xi/10$, respectively. In this
regime, as shown in  Fig. \ref{strong}(b), asymmetric coupling
strength will introduce little influence, and thus it is more robust
against coupling strength deviation than that of the small coupling
regime. With the increase (decrease) of $g_1$ from the symmetric
value of 1, the transfer time will be shortened (prolonged) and
the fidelity of the QST will be gradually increased (decreased).

\begin{figure*}[tb]\centering
\includegraphics[width=12cm]{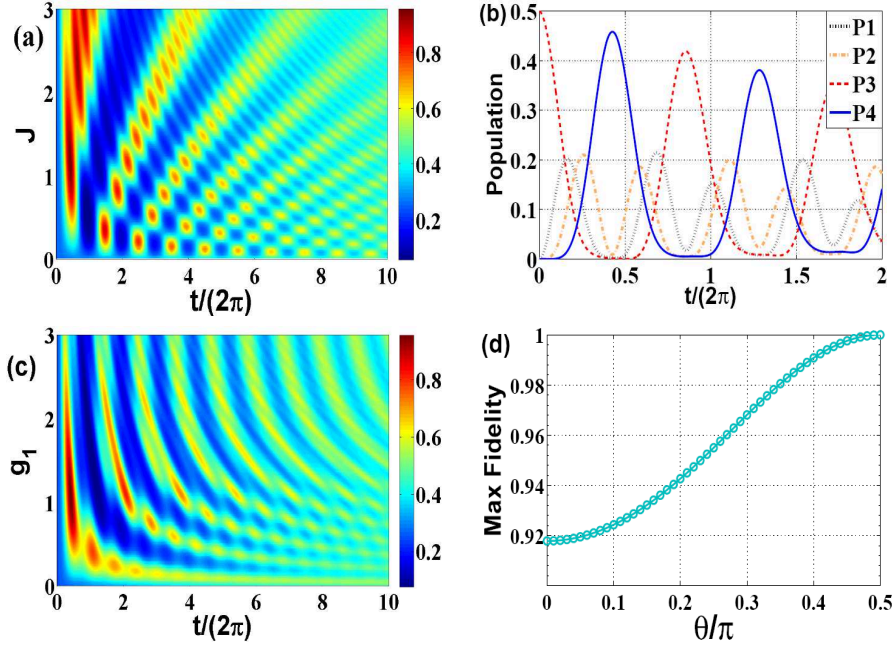}
\caption{Fidelity as a function of the reduced time of $t/(2\pi)$  in the intermediate coupling regime with $\xi=0.03$ and $\zeta=0.001$. (a).  For $\theta=\pi/4$,  $g_1=g_2=1$ and $J \in [0.01, 3]$.
(b). Time dependent populations with $J = 1.2$ in (a).
(c) For $\theta=\pi/4$,  $J=1.2$, $g_2=1$  and  $g_1 \in [0.01, 3]$. (d).  Maximum fidelity as a function of $\theta$ with $g_1=g_2=1$ and $J=1.16$.} \label{3d}
\end{figure*}

\bigskip

\noindent\textbf{The intermediate coupling regime.}
In the following, we will focus on the QST performance in the
intermediate coupling regime,  which is the best choice for QST
purpose. First,  the QST in the small coupling regime needs very
long time ($T_w \sim 1/J \gg 1/g$), and thus decoherence will
introduce huge errors. Second, the interaction between the
resonators is usually weak, or at least is the same order as the
qubit-resonator interaction $g$. Though we may tune $g$ to match
the condition of $J \gg g$, i.e., the large coupling regime, the
expense is that the effective coupling will be decreased and the
time to complete the QST will also be relatively long: $T_s \sim
1/J_{\text{eff}}=J/g^2 \gg 1/g$. This means that the time needed
for QST in the two regimes are  much longer than their respective
time scale of  $1/g$, and thus decoherence will cause intolerable
influence in both regimes. However, for the intermediate coupling
regime $J\sim g$, as the time needed for QST will be $T_i \sim 1/J
\sim 1/g$, we expect higher fidelity QST than that in the
two other regimes.

For verification purpose, we plot the fidelity of the QST process in Fig. \ref{3d}(a) with $g_2=g_1=1$, $\xi=0.03$, and $0.01\leq J\leq 3$. It is clear that
the fidelity presents regular distribution in the whole region and each
red spot corresponds to  a maximum. When $J$ increases gradually
from $0.01$ to $1.2$,  the red spot moves closer and closer to the
vertical axis. When $J\simeq 1.2$, the fidelity reaches a maximum value
with the shortest QST time. If one continues to increase $J$
towards $J > 1.2$, the red spot moves away from the vertical axis.
It means that the time needed for QST will increase gradually and
at the same time the fidelity will decease gradually. Comparing
with that in the Fig. \ref{weak} and Fig. \ref{strong}, the QST
here completes in a much shorter time (the maximum value of  the fidelity
here move towards the vertical axis), which leads to the increase
of QST efficiency and fidelity. Therefore, our numerical results
show that the QST performance in the intermediate regime can be
greatly enhanced comparing to that in the other two regimes.
In order to have an insight into the detail population changes, we
focus on the red spot closest to the vertical axis, the
populations for different states are plotted in Fig. \ref{3d}(b).
The amplitude of $P_4$ reaches the largest magnitude, implying the
initial state fully transferred from the superconducting qubit to
the NV center. Therefore, $J \sim g$ makes the
quantum bus to be a expressway and the excitation can pass through
faster and with higher fidelity.

Furthermore, we investigate the influence of asymmetric coupling
$g_2 \neq g_1$ to the QST in this regime. Figure \ref{3d}(c) shows the time-dependent population of the target state $P_4$  with $\xi=0.03$, $g_2=1$, $J=1.2$ and $0.01\leq g_1 \leq 3$. One can easily find that the amplitude of
$P_4$ enhances and rapidly moves to the left when $g_2$ increases
gradually from $0.01$. This improvement does not means the success
of faithful QST, due to the additional influence by vertical
adjacent peaks. However, within the small region of $g_1=1.1 \pm 0.1$, the
peaks are relatively far apart from each other in vertical
direction, which makes the higher fidelity QST to be realizable.
We have confirmed that one can perfectly realize the QST with
$g_1=g_2=1$ and $J\simeq1.16$, where the QST needs less time
(which is about $1/10$ and $1/5$ of the time in small and large
coupling regimes, respectively), and the QST has the highest fidelity
of $96\%$. We further investigate the $\theta$ dependence of the maximum fidelity, as shown in fig. \ref{3d}(d), which indicates that the larger  $\theta$ will lead to larger fidelity.

In conclusion, we have demonstrated that a  high fidelity QST between a superconducting qubit and   an NV center in diamond in a hybrid electromechanical system can be realized in the intermediate coupling region. We have numerically simulated the evolution of whole system and find that QST can be realized much
faster and more reliable in the intermediate coupling regime than that in the small or large coupling regimes. As the above conclusion is model independent, our result may be directly extended to similar systems consisting of coupled resonators and thus provides a promising way for realizing high fidelity QST.

\bigskip

\noindent\textbf{Method}

\noindent\textbf{Modeling of decoherence effects.}
Inevitably, the QST process will suffer from  decoherence. We
simulate the performance of our scheme under realistic conditions
by  considering the decays of the two resonators ($\kappa_a$ and
$\kappa_b$), the relaxation   ($\gamma_1$ and $\gamma_2$) and   dephasing ($\Gamma_1$ and $\Gamma_2$) of the two qubits. Under these conditions, the whole system can be
described by the Lindblad master equation
\begin{eqnarray}\label{master1}
\dot\rho = -i[H_t, \rho]
+ \frac 1 2 \left(\kappa_a \mathcal{L}(a) + \kappa_b \mathcal{L}(b)\right) \nonumber\\
 + \frac 1 2 \sum_{j=1,2} \left[\gamma_j \mathcal{L}(\sigma_j^-) + \Gamma_j \mathcal{L}(\sigma_j^z) \right],
\end{eqnarray}
where $\mathcal{L}(A)=2A\rho A^\dagger-A^\dagger A \rho -\rho A^\dagger A$ is the Lindblad operator. We next justify our chosen parameters. The superconducting qubit in a transmission-line  resonator can be protected by  the resonator, for a planar transmon qubit,  relaxation and coherence times of 44 and 20 $\mu$s are reported for a 1D resonator \cite{sq}, which leads to $\Gamma_1/2\pi =8$ kHz and $\gamma_1/2\pi = 3.5$  kHz.  The  decay rate of the 1D transmission-line resonator is around $\kappa_a/2\pi\simeq 3.5$ kHz (lifetime $\tau_a$= 45 $\mu$s \cite{tlr}). The coupling between the superconducting qubit and the transmission-line resonator is very strong and can be controlled very well, and thus we assume $g_1$ can be tuned to arbitrary value we want, which can be achieved by using a flux-biased rf SQUID to couple the  superconducting qubit to the cavity \cite{tune}. For a temperature of T=100 mK and the quality factor $Q=10^6$, the decay rate of the nanomechanical resonator is $\kappa_b/2\pi\simeq 2$ kHz
\cite{Rabl2010}, and $g_2/2\pi\simeq 115$ kHz \cite{Rabl2009}. As the  decay rates are in the same order $\Gamma_1\sim\gamma_1\sim\kappa_b\sim\kappa_a$, we treat them as a same parameter $\xi=\kappa_a$.  Hereafter, all the parameters will be normalized in unit of $g_2$, and thus  $\xi= 0.03$. The energy relaxation and coherence times of the NV center can be very long comparing with others  \cite{du}, and thus  $\gamma_2$ and $\Gamma_2$ is very small. Therefore,  we have taken these two parameters as the same in the numerical simulations, i.e., $\Gamma_2=\gamma_2=\zeta=0.001$, which corresponds to a coherence time of only 1.4 ms.

\bigskip

\noindent\textbf{Acknowledgements}\\
This work was supported by  the NFRPC (Grants No. 2013CB921804, No. 2011CB922104, and No. 2011CBA00302), the NSFC (Grants No. 11125417, No. 11104096, No. 11374117, No. 11105136, No. 61033001, and No. 61361136003), the PCSIRT (No. IRT1243), and the GRF (HKU7058/11P \& HKU7045/13P) of Hong Kong.


\begin{thebibliography}{99}

\bibitem{rjs}
Schoelkopf, R. J. \& Girvin, S. M. Wiring up quantum systems. \emph{Nature}  \textbf{451}, 664-669 (2008).

\bibitem{Balasubramanian2009}
Balasubramanian, G. \emph{et al.}
%P.~Neumann, D.~Twitchen, M.~Markham, R.~Kolesov,
%N.~Mizuochi, J.~Isoya, J.~Achard, J.~Beck, J.~Tissler, V.~Jacques, P.~R.
%Hemmer, F.~Jelezko, and J.~Wrachtrup,
Ultralong spin coherence time in isotopically engineered diamond.
\emph{Nat. Mater.} \textbf{8}, 383-387 (2009).

\bibitem{hybrid}
Xiang, Z.-L., Ashhab,  S., You,  J. Q. \& Nori, F.
Hybrid quantum circuits: Superconducting circuits interacting with other quantum systems. \emph{Rev. Mod. Phys.} \textbf{85}, 623-653 (2013).


\bibitem{Tordrup2008}
 Tordrup, K. \& M{\o}lmer, K.
Quantum computing with a single molecular   ensemble and a Cooper-pair box.
\emph{Phys. Rev. A} \textbf{77}, 020301 (2008).

\bibitem{Petrosyan2009}
Petrosyan, D., Bensky, G., Kurizki, G., Mazets, I.,~Majer, J. \& Schmiedmayer, J.
Reversible state transfer between superconducting qubits and atomic   ensembles.
\emph{Phys. Rev. A} \textbf{79}, 040304 (2009).


\bibitem{Yang2011b}
Yang, W.~L., Yin,  Z.~Q., Hu,  Y., Feng, M.  \& Du, J.~F.
High-fidelity  quantum memory using nitrogen-vacancy center ensemble for hybrid quantum computation. \emph{Phys. Rev. A }\textbf{84}, 010301 (2011).


\bibitem{Kubo2011}
Kubo, Y. \emph{et al}.
%C.~Grezes, A.~Dewes, T.~Umeda, J.~Isoya, H.~Sumiya, N.~Morishita,
%H.~Abe, S.~Onoda, T.~Ohshima, V.~Jacques, A.~Dr\'{e}au, J.~F. Roch, I.~Diniz,
%A.~Auffeves, D.~Vion, D.~Esteve, and P.~Bertet,
Hybrid quantum  circuit with a superconducting qubit coupled to a spin ensemble.
\emph{Phys. Rev.  Lett.} \textbf{107}, 220501 (2011).


\bibitem{nvs}
Marcos, D., Wubs,  M., Taylor,  J. M., Aguado,  R., Lukin, M. D. \& S{\o}rensen, A. S.  Coupling Nitrogen-vacancy centers in diamond to superconducting flux qubits.
\emph{Phys. Rev. Lett.} \textbf{105}, 210501 (2010).


\bibitem{nvs2} Zhu,  X. \emph{et al}.
%S. Saito, A. Kemp, K. Kakuyanagi, S. Karimoto, H. Nakano, W. J. Munro, Y. Tokura,
%M. S. Everitt, K. Nemoto,  M. Kasu, N. Mizuochi, and K. Semba,
Coherent coupling of a superconducting flux qubit to an electron spin ensemble in diamond. \emph{Nature} \textbf{478}, 221-224 (2011).


\bibitem{nvs3} Saito,  S.  \emph{et al}.
%X. Zhu, R. Ams\"{u}ss, Y. Matsuzaki, K. Kakuyanagi, T. Shimo-Oka, N. Mizuochi, K. Nemoto, W. J. Munro, and K. Semba,
Towards realizing a quantum memory for a superconducting qubit: Storage and retrieval of quantum states. \emph{Phys. Rev. Lett.} \textbf{111}, 107008 (2013).


\bibitem{Cirac1997}
Cirac, J.~I., Zoller,  P., Kimble,  H.~J. \& Mabuchi, H.
Quantum state  transfer and entanglement distribution among distant nodes in a quantum  network. \emph{Phys. Rev. Lett.} \textbf{78}, 3221 (1997).

\bibitem{suncp}
Sun, C. P., Wei, L. F.,  Liu, Y.-X. \& Nori, F.
Quantum transducers: Integrating transmission lines and nanomechanical resonators via charge qubits. \emph{Phys. Rev. A} \textbf{73}, 022318 (2006).

\bibitem{yin2009}
Yin, Z.-Q. \& Han,  Y.-J. Generating EPR beams in a cavity optomechanical system.
\emph{Phys. Rev. A }\textbf{79}, 024301 (2009).

\bibitem{Li2009a}
Li, P.-B., Gu, Y., Gong,  Q.-H. \& Guo, G.-C.
Quantum-information  transfer in a coupled resonator waveguide.
\emph{Phys. Rev. A} \textbf{79}, 042339  (2009).

\bibitem{sing} Singh,   S., Jing,  H., Wright,  E. M. \& Meystre, P.
Quantum-state transfer between a Bose-Einstein cond ensate and an optomechanical mirror.
\emph{Phys. Rev. A} \textbf{86}, 021801  (2012).

\bibitem{Yin2007}
 Yin, Z.-Q. \& Li, F.-L.
Multiatom and resonant interaction scheme for   quantum state transfer
and logical gates between two remote cavities via an   optical fiber.
\emph{Phys. Rev. A} \textbf{75}, 012324 (2007).

\bibitem{ks1}
Stannigel, K., Rabl, P., S{\o}rensen, A. S., Zoller,  P. \& Lukin, M. D.
Optomechanical transducers for long-distance quantum communication.
\emph{Phys. Rev. Lett.} \textbf{105}, 220501 (2010).

\bibitem{ks2} Stannigel, K., Rabl, P., S{\o}rensen, A. S., Lukin, M. D. \& Zoller, P. Optomechanical transducers for quantum-information processing.
\emph{Phys. Rev. A} \textbf{84}, 042341 (2011).


\bibitem{zhang2010a}
Zhang K. \& Li, Z.-Y.
Transfer behavior of quantum states between   atoms in photonic crystal coupled cavities. \emph{Phys. Rev. A} \textbf{81},   033843 (2010).

\bibitem{Yang1}
Yang, W.~L.,~Hu,  Y., Yin, Z.~Q., Deng,  Z.~J. \& Feng, M.
Entanglement  of nitrogen-vacancy-center ensembles using transmission line resonators and a
superconducting phase qubit.  \emph{Phys. Rev. A} \textbf{83}, 022302 (2011).

\bibitem{Yang2}
 Yang, W.~L., Yin,  Z.~Q., Xu, Z.~Y., Feng, M. \& Oh, C.~H.
Quantum  dynamics and quantum state transfer between separated nitrogen-vacancy
centers embedded in photonic crystal cavities. \emph{Phys. Rev. A} \textbf{84},   043849 (2011).


\bibitem{Rabl2009} Rabl,  P., Cappellaro,  P., Gurudev Dutt, M. V., Jiang,  L., Maze, J. R. \& Lukin, M. D.
Strong magnetic coupling between an electronic spin qubit and a mechanical resonator.
\emph{Phys. Rev. B} \textbf{79}, 041302 (2009).

\bibitem{Rabl2010}
Rabl,  P., Kolkowitz, S.~J., Koppens, F.~H.~L., Harris, J.~G.~E., Zoller, P.  \& Lukin, M.~D. A quantum spin transducer based on  nanoelectromechanical resonator arrays. \emph{Nat. Phys.} \textbf{6}, 602-608 (2010).


\bibitem{Arcizet2011}
Arcizet, O., Jacques, V., Siria, A., Poncharal, P., Vincent, P. \& Seidelin, S.
A single nitrogen-vacancy defect coupled to a nanomechanical   oscillator.
\emph{Nat. Phys.} \textbf{7}, 879-883 (2011).


\bibitem{em} Teufel, J. D., Li, D., Allman, M. S., Cicak, K., Sirois, A. J., Whittaker, J. D.. \& Simmonds, R.W.
Circuit cavity electromechanics in the strong-coupling regime.
\emph{Nature} \textbf{471}, 204-208 (2011).


\bibitem{Zhou2013}
Zhou, X., Hocke,  F., Schliesser, A., Marx, A., Huebl, H., Gross, R. \&   Kippenberg, T.~J.
Slowing, advancing and switching of microwave signals   using circuit nanoelectromechanics. \emph{Nat. Phys.} \textbf{9}, 179-184 (2013).

\bibitem{Pirkkalainen2013}
 Pirkkalainen, J.-M., Cho, S.~U., Li, J., Paraoanu, G.~S., Hakonen, P.~J.  \&   Sillanp\"{a}\"{a}, M.~A.
Hybrid circuit cavity quantum electrodynamics   with a micromechanical resonator.
\emph{Nature} \textbf{494}, 211-215 (2013).

\bibitem{Palomaki2013}
Palomaki, T.~A., Harlow, J., Teufel,  J.~D., Simmonds, R.~W. \& Lehnert, K.~W.
Coherent state transfer between itinerant microwave fields and a   mechanical oscillator. \emph{Nature}  \textbf{495}, 210-214 (2013).


\bibitem{em1} Vitali, D., Tombesi, P., Woolley, M. J., Doherty, A. C. \& Milburn G. J. Entangling a nanomechanical resonator and a superconducting microwave cavity. \emph{Phys. Rev. A} \textbf{76}, 042336 (2007).

\bibitem{em2} Barzanjeh, Sh., Vitali, D., Tombesi, P.  \& Milburn, G. J.
Entangling optical and microwave cavity modes by means of a nanomechanical resonator. \emph{Phys. Rev. A} \textbf{84}, 042342 (2011).

\bibitem{em3} Li, P.-B., Gao, S.-Y. \& Li, F.-L.
Robust continuous-variable entanglement of microwave photons with cavity electromechanics. \emph{Phys. Rev. A} \textbf{88}, 043802 (2013)


\bibitem{blias} Blais, A., Huang,  R.-S., Wallraff, A., Girvin, S. M. \& Schoelkopf, R. J.
Cavity quantum electrodynamics for superconducting electrical circuits: An architecture for quantum computation.
\emph{Phys. Rev. A} \textbf{69}, 062320 (2004).

\bibitem{zhu}  Zhu, S.-L., Wang, Z. D. \& Yang, K. Quantum-information processing using Josephson junctions coupled through cavities. Phys. Rev. A \textbf{68}, 034303 (2003).

\bibitem{du} Shi, F. \emph{et al}.
% X. Rong, N. Xu, Y. Wang, J. Wu, B. Chong, X. Peng, J. Kniepert, R.-S. Schoenfeld, W. Harneit, M. Feng, and J. Du,
Room-temperature implementation of the Deutsch-Jozsa algorithm with a single electronic spin in diamond. \emph{Phys. Rev. Lett}. \textbf{105}, 040504 (2010).

\bibitem{enhance} Heikkil\"{a}, T. T., Massel, F., Tuorila, J., Khan,  R. \& Sillanp\"{a}\"{a} M. A. Enhancing optomechanical coupling via the Josephson effect. \emph{Phys. Rev. Lett}. \textbf{112}, 203603 (2014).



\bibitem{strain} Kepesidis, K. V., Bennett, S. D., Portolan, S., Lukin, M. D. \& Rabl P. Phonon cooling and lasing with nitrogen-vacancy centers in diamond. \emph{Phys. Rev. B} \textbf{88}, 064105   (2013).

\bibitem{sq} Barends, R. \emph{et al}.
%J. Kelly, A. Megrant, D. Sank, E. Jeffrey, Y. Chen, Y. Yin, B. Chiaro, J. Mutus,
%C. Neill, P. O'Malley, P. Roushan, J. Wenner, T. C. White, A. N. Cleland, and J. M. Martinis,
Coherent Josephson qubit suitable for scalable quantum integrated circuits. \emph{Phys. Rev. Lett}. \textbf{111}, 080502 (2013).

\bibitem{tlr}
Megrant, A. \emph{et al}.
%C. Neill, R. Barends, B. Chiaro, Y. Chen, L. Feigl, J. Kelly, E. Lucero, M. Mariantoni, P. J. J. O'Malley,
%D. Sank, A. Vainsencher, J. Wenner, T. C. White, Y. Yin, J. Zhao, C. J. Palmstr{\o}m, J. M. Martinis, and A. N. Cleland,
Planar superconducting resonators with internal quality factors above one million.
\emph{Appl. Phys. Lett.} \textbf{100}, 113510 (2012).

\bibitem{tune} Allman, M. S., Altomare, F.,  Whittaker, J. D.,  Cicak, K., Li, D., Sirois, A., Strong, J., Teufel, J. D. \& Simmonds, R.W. rf-SQUID-mediated coherent tunable coupling between a superconducting phase qubit and a lumped-element resonator. Phys. Rev. Lett. 104, 177004 (2010).


\end{thebibliography}
\end{document}